\documentclass[a4paper,12pt]{article}
\usepackage[utf8]{inputenc}
\usepackage{cancel}
\usepackage{ulem}
\usepackage{amsfonts}
\usepackage{amssymb}
\usepackage{graphicx}
\usepackage{amsmath}
\usepackage{enumerate}
\usepackage{mathtools}
\usepackage{subfig}
\usepackage{color}
\usepackage{tikz}\usetikzlibrary{calc}
\usepackage{float}
\usepackage{here}
\usepackage{cite}
\usepackage{mathrsfs}
\usepackage{float,epsfig}
\usepackage{dcolumn}

\usepackage{graphicx}
\usepackage{bm}
\usepackage{amsmath,amssymb,amsthm}
\usepackage[colorlinks=true,linkcolor=blue,citecolor=red]{hyperref}
\newcommand{\be}{\begin{equation}}
\newcommand{\ee}{\end{equation}}
\newcommand{\bea}{\setlength\arraycolsep{2pt} \begin{eqnarray}}
\newcommand{\eea}{\end{eqnarray}}

\def\0{{\sst{(0)}}}
\def\1{{\sst{(1)}}}
\def\2{{\sst{(2)}}}
\def\3{{\sst{(3)}}}
\def\4{{\sst{(4)}}}
\def\5{{\sst{(5)}}}
\def\6{{\sst{(6)}}}
\def\7{{\sst{(7)}}}
\def\8{{\sst{(8)}}}
\def\sst#1{{\scriptscriptstyle #1}}

\makeatletter \@addtoreset{equation}{section}

\usepackage{multirow}
\definecolor{lime}{HTML}{A6CE39}
\newcommand{\orcidicon}{%
    \begin{tikzpicture}
    \draw[lime, fill=lime] (0,0)
        circle [radius=0.16]
        node[white] {{\fontfamily{qag}\selectfont \tiny ID}};
    \draw[white, fill=white] (-0.0625,0.095)
        circle [radius=0.007];
    \end{tikzpicture}   \hspace{-2mm}
}
\newcommand\orcidAdil{{\href{https://orcid.org/0000-0001-7623-5541}{\orcidicon}}}
\newcommand\orcidHajar{{\href{https://orcid.org/0000-0001-9510-4248}{\orcidicon}}}              
\newcommand\orcidHasan{{\href{https://orcid.org/0000-0001-7408-0910}{\orcidicon}}}
\newcommand\orcidMohamed{{\href{https://orcid.org/0000-0003-1185-0062}{\orcidicon}}}

\begin{document}
%

\title{\normalsize
{\bf \Large	  Light  Deflection Angle by    Superentropic   Black Holes  }}
\author{ \small   A. Belhaj\orcidAdil\!\! $^{1}$\footnote{a-belhaj@um5r.ac.ma},  H. Belmahi\orcidHajar\!\!$^{1}$\footnote{hajar\_belmahi@um5.ac.ma},  M. Benali\orcidMohamed\!\!$^{1}$\footnote{mohamed\_benali4@um5.ac.ma}, H. El Moumni\orcidHasan\!\!$^{2}$\thanks{h.elmoumni@uiz.ac.ma}
	\footnote{ Authors in alphabetical order.}
	\hspace*{-8pt} \\
	{\small $^1$ D\'{e}partement de Physique, Equipe des Sciences de la mati\`ere et du rayonnement, ESMaR}\\
{\small   Facult\'e des Sciences, Universit\'e Mohammed V de Rabat, Rabat,  Morocco} \\
	{\small $^{2}$  EPTHE, D\'{e}partement de Physique, Facult\'e des Sciences,   Universit\'e Ibn Zohr, Agadir, Morocco} 
} \maketitle

 \maketitle
\begin{abstract}
	Motivated  by recent works on light deflection and shadow behaviors on AdS geometries \cite{MAN,B110},  we investigate    the deflection angle  of light rays   by  superentropic black holes.   Taking    appropriate approximations,   we  first obtain the  involved expression.     For large values of the impact parameter,   we  get a specific  value    being  zero for  ordinary black holes without AdS backgrounds. 
	Then, we  examine  and analyze such an optical quantity by providing graphical discussions in terms of  a bounded region of  the moduli space required by superentropic black hole conditions.    Concretely,  we    study  the deflection  angle aspects  by  varying the   black hole mass being fixed in the previous  findings. 
		{\noindent}
		
	\end{abstract}

\newpage


\newpage

\section{Introduction}
Recently, many efforts have been devoted to study  the black hole physic.  Precisely, the analogy between  the associated mechanics and  the thermodynamic  laws pushes one  to consider the black hole as a thermodynamic object  \cite{SWH,F,F1,E}. This has become more interesting by identifying the cosmological constant with  its pressure. Concretely, the stability, the phase transitions, and the  P-V criticality  of various black holes  have been investigated.   For  Anti de Sitter (AdS) black holes,   many  similarities with van der Waals fluids have been  established\cite{E1,12,f2,D}.

More recently,     the  pulsar SGR J174-2900 near  supermassive black holes  SgrA* has  been explored  and investigated  providing     data on    the  horizon and the  horizonless  of events around   compact objects  \cite{BW1,BW2}. A special emphasis  has been put on superentropic  black holes as  fascinate  descriptions  with non compact horizon topologies  having  an entropy of maximum bounds  \cite{BW3,BW4,BW44,BW444}.    Relaying   on certain   limits, the superentropic  black holes      have been  approached by considering  ultra-spinning	limits of the  Kerr-Newman-AdS solution \cite{BW5}.   Indeed, the thermodynamic   of such black holes has  been investigated    by establishing     a  mapping  with ordinary  solutions  based on  ultra-spinning  appropriate approximations\cite{BW6,BW7,BW70,BW71}. 

After  many years of investigations, the image of the supermassive black hole provided by  the Event  Horizon Telescope (ETH) and the detection of gravitational waves by LiGO    have  brought    a greatest advancement in  modern physics \cite{BW8,BW9,A2,A3,A4}.   These  have encouraged activities concerning the optical aspects of    black hole  objects. The  obtained behaviors have unveiled more data which could be exploited to understand   the  corresponding  interesting   issue. Concretely, the shadow of  various  black holes has been engineered,   in the   celestial two  dimensional space,  using null  geodesic equations  based on  the Hamilton-Jacobi approach.   The corresponding physical  information have been  encoded in the shadow radius and the distortion parameter. These two  geometrical parameters control the size and the shape of the involved shadows, respectively. For  non-rotating  black holes,  for instance, it has been revealed  that  the shadows  involve a circular  configuration shape.  The later has been distorted by introducing the rotating parameter to provide non-trivial geometries including  either the D-shape, or cardioid forms \cite{H, B12,BC}.    Moreover,    a particular interest has been put on  the deflection angle behaviors   of light rays    using analytic and numerical  methods and approaches.  Precisely, Gibbons and  Werner brought    a direct way to  compute   such an optical quantity  using the Gauss-Bonnet theorem  applied to a spacial background described by the optic metric functions \cite{BW10,BW11,BW12,BW13,BW15,BW16,BW17,BW18,BW19,BW20,BW21}.  Moreover, it has been suggested  an alternative  approach  based on  an  elliptic integral formalism using Weierstrass elliptic functions\cite{BW22}.

In this work, we compute and analyze the light deflection angle by superentropic black holes in four dimensions. Up to certain assumptions and appropriate approximations, we first find the associated expression.       For large values of the impact parameter,   we   obtain a specific  value    being  zero for  ordinary black holes without AdS backgrounds. 
Then, we  examine  and  discuss  such an optical quantity by providing graphic representations  in terms of the involved parameters.    Motivated by the mass constraint required by  superentropic black hole conditions,  we    discuss the deflection  angle aspect by varying the   black hole mass being fixed in the previous  investigations.

The organization of this work is as follows. Section 2 concerns the needed formalism to compute the involved optical quantity. Section 3  brings  the obtained expression. Section 4 is developed  to graphic representations. The last section  provides concluding  remarks.

 \section{ Deflection angle computations}
 To compute the deflection angle of light rays  around four dimensional  black holes,  several methods have been explored.  They  have been extensively investigated to approach a large class  of solutions. One of them  has been relied on  the geodesic equations of  the massless   particle motion. This road, however,  provides   solutions using non-trivial  elliptic functions. An alternative  method, which will be  used   in the present work, exploits the Gauss-Bonnet formalism results  based on  optic metric calculations   \cite{BW23,BW24,BW15,BW27}.
Placing  the observer and the source  at finite distance in the equatorial plane, the deflection angle  can be derived  using the following equation
 \begin{equation}
 \Theta=\Psi_{R}-\Psi_{S}+\phi_{SR}.
 \label{a1}
 \end{equation}
In this equation, $\Psi_{R}$ and $\Psi_{S}$  indicate angles between the light ray and the radial direction at the observer and  the source position,  respectively.  It is denoted that  $\phi_{SR}$  represents the longitude separation angle specified later on \cite{BW24}. These angles  are linked to the components of  the unit tangential vector along the light $e^i$ defined  by
\begin{equation}
\label{L2}
(e^r,e^\theta,e^\phi)=\epsilon\left(\frac{dr}{d\phi},0,1\right).
\end{equation}
where $\epsilon$ can be obtained from the metric.
Using the following   metric relation 
\begin{equation}
ds^2=-A(r,\theta)dt^2+B(r,\theta)dr^2+C(r,\theta)d\theta^2+D(r,\theta)d\phi^2-2H(r,\theta)dt d\phi,
\end{equation}
$\epsilon$ can be determined as as  a function of the metric parameters. Indeed, it is given by  
\begin{equation}
\label{L3}
\epsilon=\frac{A(r)D(r)+H^2(r)}{A(r)(H(r)+A(r)b)}
\end{equation}
where $b$ is the impact parameter specified later on. In the equatorial plane $(\theta=\pi/2)$ at  a constant $t$ of the space-time metric,  it is possible to   consider   a  2-dimensional curved space   defined by the following line element 
\begin{equation}
dl^2\equiv \gamma_{ij}dx^{i}dx^{j}
\label{labe}
\end{equation}
where $\gamma_{ij}$ represents  a spatial metric\cite{BW13,BW15,BW23}. To obtain  the  angle measured from the radial direction,   one  should use  the expression of   the unit basis vector and the spatial metric  as following
\begin{equation}
\label{L1a}
\cos\Psi\equiv \gamma_{ij}e^iR^j,
\end{equation}
where  the components of the radial vector $R^j$ are $(\frac{1}{\sqrt{\gamma_{rr}}},0,0)$    taken at    the particular angle $\theta=\pi/2$.  Using Eq.(\ref{L2}) and Eq.(\ref{L1a}),     the $\sin\Psi$ expression is found to be 
\begin{equation}
\label{G111}
\sin\Psi=\frac{H(r)+A(r)b}{\sqrt{A(r)D(r)+H^2(r)}}.
\end{equation}
Having presented shortly the  formalism needed  to compute  the deflection angle of lights,  we move  to  superentropic   black hole applications. 
\section{Deflection angle of superentropic  black holes}
Motivated by  a recent work on   black hole shadows,  we  would like to investigate  the deflection angle of lights by   special black hole  solutions called  superentropic  black holes. These solutions have  been proposed as    the Kerr-Neuman-AdS black hole limits.  Before calculating  such a  deflection  angle of light rays,  we first  present  relevant   physical and mathematical  elements corresponding to these solutions.    Following  \cite{B110,BW3,BW5},  the line element of  the associated  metric  reads as
\begin{eqnarray}
ds^{2}=&-&\frac{\Delta_{r}}{\Sigma} \left(dt - {\ell} \sin^2\theta d\phi \right)^2 +\Sigma  \left(\frac{dr^2}{\Delta_{r}}
+\frac{d\theta^2}{\sin^2\theta} \right)\\
\nonumber
&+& \frac{ \sin^4\theta}{\Sigma} \left( \ell dt -{(r^2+\ell^2)}d\phi \right)^2,
\end{eqnarray}
where  $\ell$ is a  cosmological parameter.   The involved metric terms 
\begin{equation}
\Sigma=r^2+\ell^2\cos^2\theta, \hspace{0.5 cm} \Delta_r=(\ell+\frac{r^2}{\ell})^2-2mr+q^2.
\end{equation}
It is noted that  $m$ and $q$ are the mass and the charge parameters, respectively. In this solution,   the usual  local  coordinate $\phi$,   being a  noncompact direction,  should be compactified   via the relation  $\phi \sim \phi + \alpha $,  where  $\alpha$ is a   dimensionless parameter controlling the associated size. The later  has been  identified with a  new chemical potential   as proposed in \cite{BW5}.  The solution of the equation $\Delta_r=0$ provides a large root $r_+$  which reveals  the existence of the black hole horizon.   A close examination shows  that  this solution    requires a  constraint on  the mass parameter. The calculation   has given     
\begin{equation}
\label{ES}
m\geqslant 2r_c(\frac{r_c^2}{\ell^2}+1),
\end{equation}
where one has a critical radius via the relations $r_c^2=\frac{\ell^2}{3}\left[(4+\frac{3}{\ell^2}q^2)^{1/2}-1\right]$.  This constraint  provides  a relationship between  $m$,  $\ell$ and $q$  black hole parameters. It turns  out that the thermodynamics of such black hole solutions  has   been investigated in  many places including in \cite{BW6,BW7}.  In particular,  certain quantities  have been approached in terms of usual  charged and rotating black holes.    In  the present  investigation,  we attempt to  complete   the associated optical  behaviors.  This goes beyond the previous  works  where  the mass   parameter has been fixed \cite{B12,BW21}.    Here,  however,  this black hole parameter  will be varied  up to the previous constraint.    To inspect   the deflection angle behaviors of such  four dimensional black holes,   we present  the associated backgrounds from  the  orbit equation on the equatorial plane.  More precisely, it  is expressed as follows 
\begin{equation}
\label{eq}
 \left( \frac{dr}{d\phi}\right) ^{2}=\frac{ \left(A(r)D(r)+H^2(r)\right)  \left(D(r)-2H(r)b-A(r)b^2\right) }{B(r) \left(H(r)+A(r)b\right)^2},
\end{equation}
where   $b$ is  the impact parameter  given by  
\begin{equation}
b=\frac{\mid L \mid}{E}=\frac{D(r)\frac{d\phi}{dt}-H(r)}{H(r)\frac{d\phi}{dt}+A(r)}.
\end{equation}
The involved  radial functions read as 
\begin{eqnarray}
\label{eq1}
A(r)&=&\frac{\Delta_r-\ell^2}{\Sigma} \hspace{2 cm} D(r)=\frac{(r^2+\ell^2)\ell-\Delta_r\ell}{\Sigma}\\
\label{eq2}
B(r)&=&\frac{\Sigma}{\Delta_r}  \hspace{2.8 cm} H(r)=\frac{(r^2+\ell^2)^2-\Delta_r\ell^2}{\Sigma}.
\end{eqnarray}
Using  the expressions given  in Eq.(\ref{eq1})  and  Eq.(\ref{eq2}) and taking  $u=\frac{1}{r}$,  Eq.(\ref{eq}) becomes
\begin{equation}
\label{F1}
 \left( \frac{du}{d\phi}\right) ^{2}=\frac{\Big(1+\ell^4u^4+\alpha\Big)^2\Big(\ell^2(\alpha-2\ell^2u^2)-2b\ell(1-\ell^2u^2+\alpha)+b^2(1+\alpha)\Big)}{\ell^2\Big(\ell(1-\ell^2u^2+\alpha)-b(1+\alpha)\Big)^2},
 \end{equation}
where  one has 
\begin{equation}
\label{ }
\alpha=\ell^2u^2(2-2mu+q^2u^2).
\end{equation}
To simplify the computations,  certain approximations should be exploited.     Considering  the following order $\mathcal{O}(m^2,q^3,\ell^2)$, the orbit equation   can take the following form 
\begin{equation}
\label{eq6}
   \left( \frac{du}{d\phi}\right) ^{2}=\frac{1}{b^2}+\frac{2 \ell}{b^3}-2 u^2+2 m u^3-q^2 u^4.
\end{equation}
The calculation of the deflection angle of light rays    needs  the explicit   expression of  $\phi_{RS}$.  Indeed,  it is given by 
\begin{equation}
\phi_{RS} = \int^R_S d\phi= \int^{u_0}_{u_S}\frac{1}{\sqrt{G(u)}}du +\int^{u_0}_{u_R}\frac{1}{\sqrt{G(u)}}du , 
\end{equation}
where  one has used  $G(u) \equiv ( \frac{du}{d\phi}) ^{2} $.    It is recalled that  $ u_S $ and $  u_R$  represent the inverse of the source and the observer distance from the black hole. Moreover,  $  u_0$  denotes  the inverse of  the closest approach $r_{0}$.

 To write down the desired expression, we  take  weak field and slow rotation approximations.   For simplicity reasons, we consider the  following order  $\mathcal{O}({m^2,\ell^2,q^2)}.$   Eq.(\ref{F1}) provides a  relation between the impact parameter $b$ and  $u_0$   which reads as 
\begin{equation}
\label{A5}
b =\ell+\frac{1}{ \sqrt{2}}\left(\frac{m}{2}+\frac{1}{u_0}\right)
\end{equation}
where $u_0$ solves   the constraint  $G(u)=0$. Combining  the above equations, we obtain 
\begin{align}
\phi_{RS} 
=&
\int^{u_0}_{u_S}\left(\frac{1}{\sqrt{2}\sqrt{{u_0}^2-u^2}}
+m\frac{ \left(u^2+u u_0+u_0^2\right)}{2 \sqrt{2} (u+u_0) \sqrt{u_0^2-u^2}}
{}-q^2\frac{ {u_0} \left(u^2+u {u_0}+2 u_0^2\right)}{4 \sqrt{2} (u+{u_0}) \sqrt{u0^2-u^2}}
\right) du 
\notag\\
&
+\int^{u_0}_{u_R}\left(\frac{1}{\sqrt{2}\sqrt{{u_0}^2-u^2}}
+m\frac{ \left(u^2+u u_0+u_0^2\right)}{2 \sqrt{2} (u+u_0) \sqrt{u_0^2-u^2}}
{}-q^2\frac{ {u_0} \left(u^2+u {u_0}+2 u_0^2\right)}{4 \sqrt{2} (u+{u_0}) \sqrt{u0^2-u^2}}
\right) du 
\notag\\ 
&+ \mathcal{O}({m^2,\ell^2,q^3}).
\end{align}
The computing integrals give 
\begin{align}
\phi_{RS} 
=&\left(\frac{1}{\sqrt{2}}\left(\frac{\pi}{2}-\arcsin\Big(\frac{u_S}{u_0}\Big)\right)
+\frac{m}{2\sqrt{2}}\frac{(2u_0+u_S)\sqrt{{u_0}^2-{u_S}^2}}{u_0+u_R}-q^2\frac{ u_0 (u_s+3 u_0) \sqrt{u_0^2-u_s^2}}{4 \sqrt{2} (u_s+u_0)}
\right) \notag\\
&+\left(\frac{1}{\sqrt{2}}\left(\frac{\pi}{2}-\arcsin\Big(\frac{u_R}{u_0}\Big)\right)
+\frac{m}{2\sqrt{2}}\frac{(2u_0+u_R)\sqrt{{u_0}^2-{u_R}^2}}{u_0+u_R}-q^2\frac{u_0 (u_r+3 u_0) \sqrt{u_0^2-u_r^2}}{4 \sqrt{2} (u_r+u_0)}
\right)
\notag\\
& +\mathcal{O}({m^2,\ell^2,q^3}) .
\label{G1}
\end{align}
It is remarked that   Eq.(\ref{G1}) is expressed as a  function of $u_0$. To get $\phi_{RS}$ in term of  the impact parameter $b$,  however, we should exploit  Eq.(\ref{A5}).   Indeed, the  calculations   lead to 
\begin{equation}
\label{EQ1}
\begin{split}
\phi_{RS}&=\frac{1 }{\sqrt{2}}\left({\pi }-\arcsin (\sqrt{2} \,b u_r)-\arcsin (\sqrt{2} \,b u_s)\right)\\
&+m\left(\frac{1-b^2 u_r^2}{2 b \sqrt{1-2 b^2 u_r^2}}+\frac{1-b^2 u_s^2}{2 b \sqrt{1-2 b^2 u_s^2}}\right)+\ell\left(\frac{{u_r}}{ \sqrt{1-2b^2 u_r^2}}+\frac{{u_s}}{ \sqrt{1-2b^2 u_s^2}}\right)\\
&+m\,\ell\left(\frac{2 \sqrt{2} b u_r+1+b^2u_r^2}{b^2 \left(2 b u_r+\sqrt{2}\right)^2 \sqrt{1-2 b^2 u_r^2}}+\frac{2 \sqrt{2} b u_s+1+b^2u_s^2}{b^2 \left(2 b u_s+\sqrt{2}\right)^2 \sqrt{1-2 b^2 u_s^2}}\right)\\
&-q^2\left(\frac{\left(\sqrt{2} b u_r+3\right) \sqrt{1-2 b^2 u_r^2}}{8 b^2 \left(2 b u_r+\sqrt{2}\right)}+\frac{\left(\sqrt{2} b u_s+3\right) \sqrt{1-2 b^2 u_s^2}}{8 b^2 \left(2 b u_s+\sqrt{2}\right)}\right)\\
&-mq^2\left(\frac{5 \sqrt{2} b u_r+3}{8 b^3 \left(2 b u_r+\sqrt{2}\right)^2 \sqrt{1-2 b^2 u_r^2}}+\frac{5 \sqrt{2} b u_s+3}{8 b^3 \left(2 b u_s+\sqrt{2}\right)^2 \sqrt{1-2 b^2 u_s^2}}\right)\\
&-\ell q^2\left(\frac{10 b u_r+3 \sqrt{2}}{4 b^3 \left(2 b u_r+\sqrt{2}\right)^2 \sqrt{1-2 b^2 u_r^2}} +\frac{10 b u_s+3 \sqrt{2}}{4 b^3 \left(2 b u_s+\sqrt{2}\right)^2 \sqrt{1-2 b^2 u_s^2}} \right)\\
&-m\ell q^2\left(\frac{50 b u_r+9 \sqrt{2}}{8 b^4 \left(2 b u_r+\sqrt{2}\right)^3 \left(1-2 b^2 u_r^2\right)^{3/2}}+\frac{50 b u_s+9 \sqrt{2}}{8 b^4 \left(2 b u_s+\sqrt{2}\right)^3 \left(1-2 b^2 u_s^2\right)^{3/2}}\right)\\
&+\mathcal{O}({m^2,\ell^2,q^3,mu_r^3,mu_s^3,q^3u_r^3,q^2u_s^2}).
\end{split}
\end{equation}
To  obtain  the   expression that we are after,  Eq.(\ref{G111}) should be used.    Up to certain approximations,      we  obtain  the  following relation 
\begin{equation}
\label{ }
\sin\Psi=b \ell u^2-b \ell m u^3+\frac{1}{2} b \ell q^2 u^4-1+\mathcal{O}({m^2,\ell^2,q^3}).
\end{equation}
This  gives 
\begin{equation}
\label{EQ2}
\Psi_R-\Psi_S= \sqrt{2b \ell}u_r + \sqrt{2b \ell}u_s -\frac{m  \sqrt{b \ell}}{\sqrt{2}}u_r^2-\frac{m  \sqrt{b \ell}}{\sqrt{2}}u_s^2-{2\pi}+\mathcal{O}({m^2,\ell^2,q^3,q^3u_r^3,q^3u_s^3}).
\end{equation}
It has been observed that  the divergence  terms linked to $u_r\to 0$ and $u_s\to 0$  have  been removed  in the Eq.(\ref{EQ1}) and Eq.(\ref{EQ2}).  With these limits, we can  get the corresponding deflection angle  of light rays  associated with the present  black hole solutions.    Performing certain   computations    and  using periodic conditions,     we    obtain the  following expression
\begin{equation}
\label{ }
\begin{split}
\Theta=&\frac{m}{b}+\frac{\ell m}{b^2}-\frac{3 q^2}{4 \sqrt{2} b^2}-\frac{3 \ell q^2}{2 \sqrt{2} b^3}-\frac{3  m q}{8 b^3}\\
&-\frac{9 \ell m q^2}{8 b^4}+\frac{\pi }{\sqrt{2}}+\mathcal{O}({m^2,\ell^2,q^3,mu_r^3,mu_s^3,q^3u_r^3,q^2u_s^2}).
\end{split}
\end{equation}
This  light angle deflection  of  superentropic  black hole solutions is given in terms of $m$, $\ell$ and $q$  parameters. Such parameters will be used to  discuss and illustrate the  associated behaviors.  A particular focus will be on the mass parameter  variation being fixed in the previous investigations.
\section{ Graphical discussions}
\begin{figure}[ht!]
		\begin{center}
		\centering
			\begin{tabbing}
			\centering
			\hspace{8.6cm}\=\kill
			\includegraphics[scale=.5]{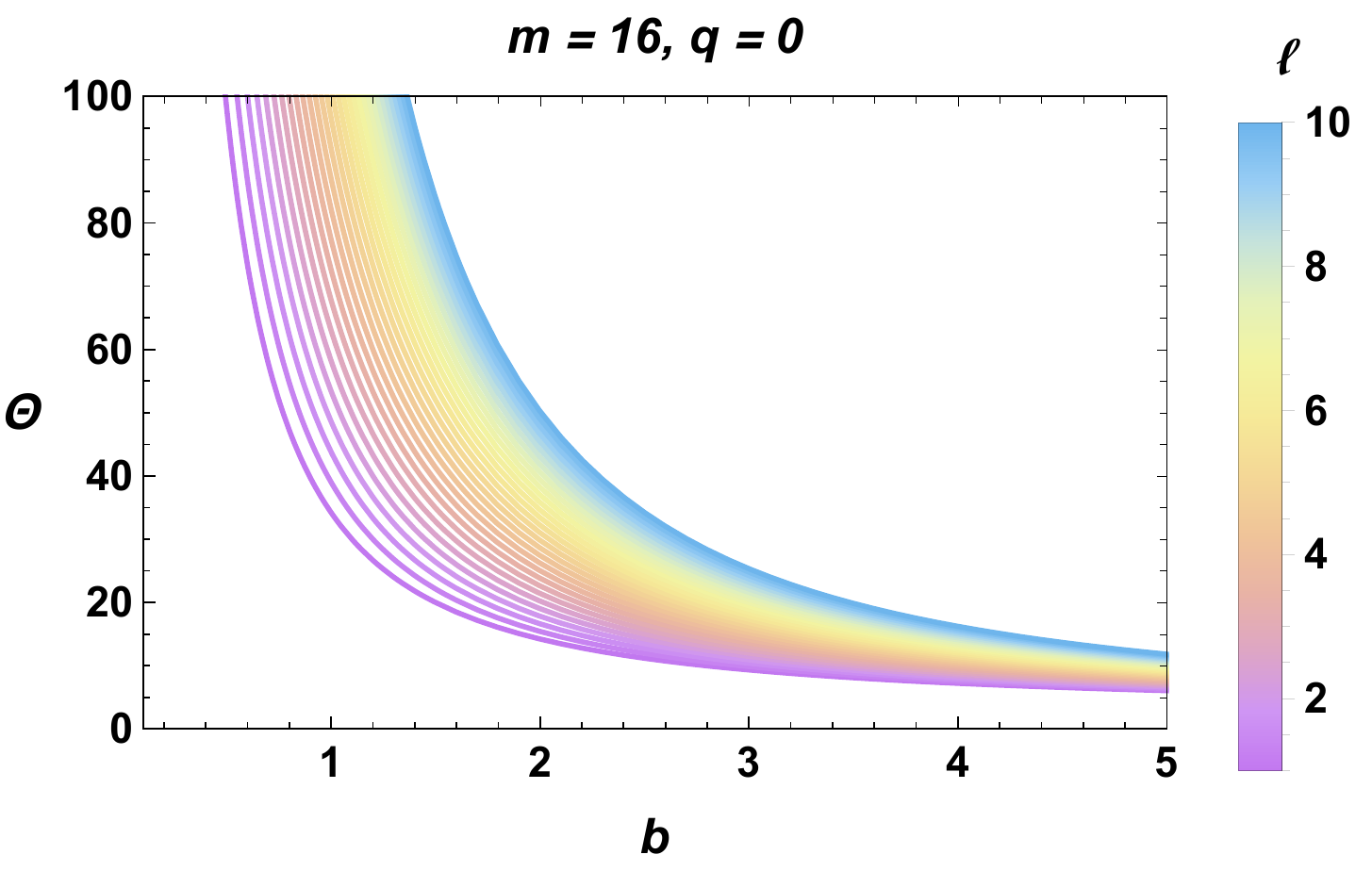} \>
			\includegraphics[scale=.5]{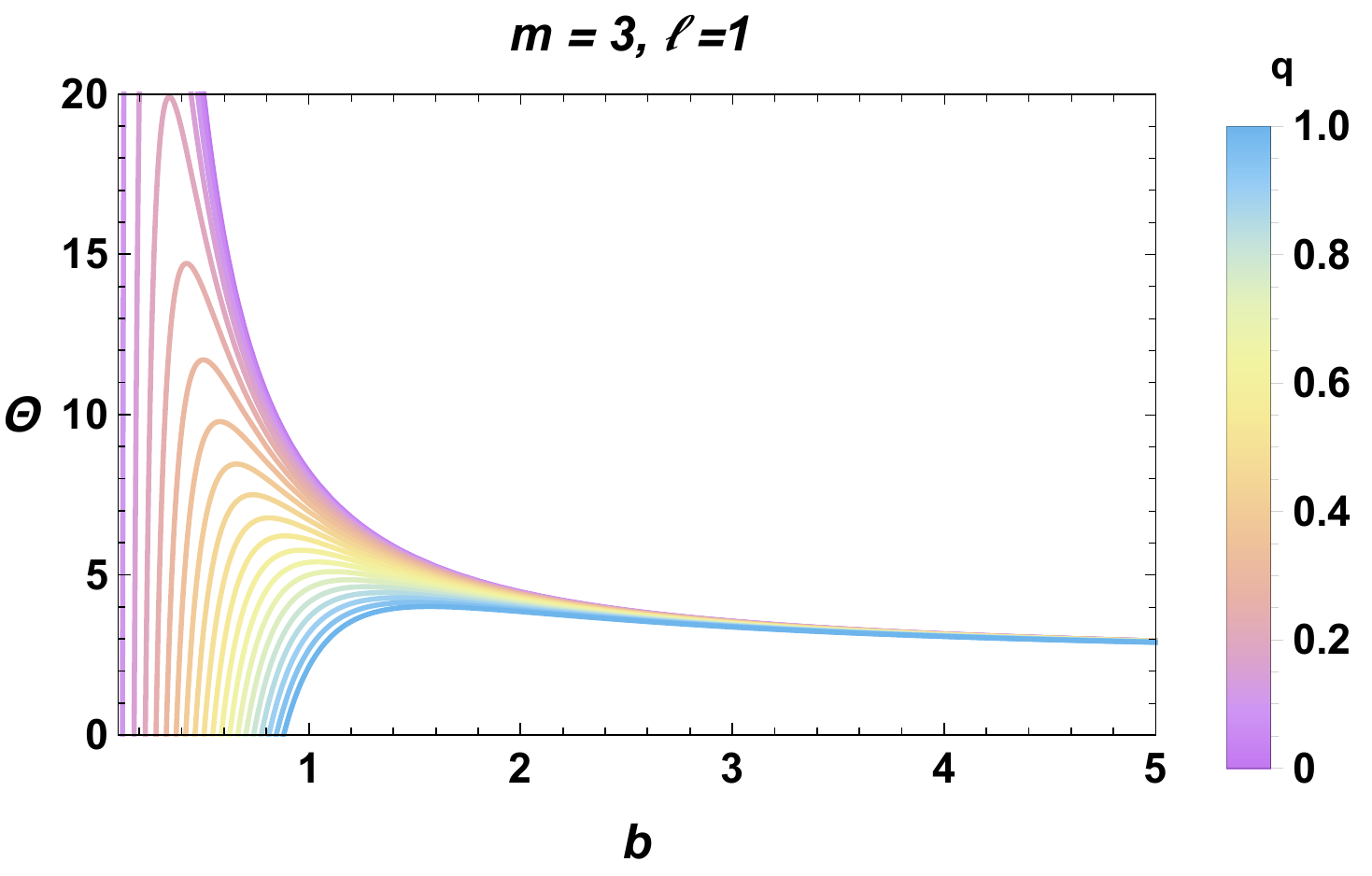} \\
					   \end{tabbing}
					   \vspace{- 1 cm}
\caption{{\it \footnotesize The deflection angle  behaviors by varying  the  impact parameter  $b$  with  different values of the relevant parameters :  the mass $m$, the AdS radius $\ell$ and   the charge $q$.
}.}
\label{shfa1}
\end{center}
\end{figure}

In this section,  we analyze the effect of  the above mentioned parameters on the deflection angle of light rays  by superentropic  black holes. In particular,  we discuss two situations  relying on the mass parameter variations. In the case of the  first situation associated with  a fixed mass parameter,  the  variation  behaviors  are illustrated in Fig.(\ref{shfa1}) for different  values 
 of the charge and the AdS length.
The left plot of this figure reveals the aspects of  the deflection angle by varying the impact parameter $b$ for different values of $\ell$ corresponding to  a neutral  solution.  The deflection angle increases  by increasing  $\ell$.   For large values of $b$, the angle deflection goes to a particular value $\frac{\pi}{\sqrt{2}}$.     Fixing the AdS length size,  the   right plot  shows  the  deflection angle behaviors  by varying the impact parameter $b$ for different   charge values of $q$.  For small values of $b$ and small  values of $q$,  the defection angle does not bring consistent acceptable behaviors.  For large values of $q$, however,   the angle deflection  keeps the  same previous variations.  For large values of $b$, this optical angle goes to $\frac{\pi}{\sqrt{2}}$.  It has been remarked that the charge can play a relevant optical role     for  such black hole solutions.

\begin{figure}[ht!]
		\begin{center}
		\centering
			\begin{tabbing}
			\centering
			\hspace{8.6cm}\=\kill
			\includegraphics[scale=.5]{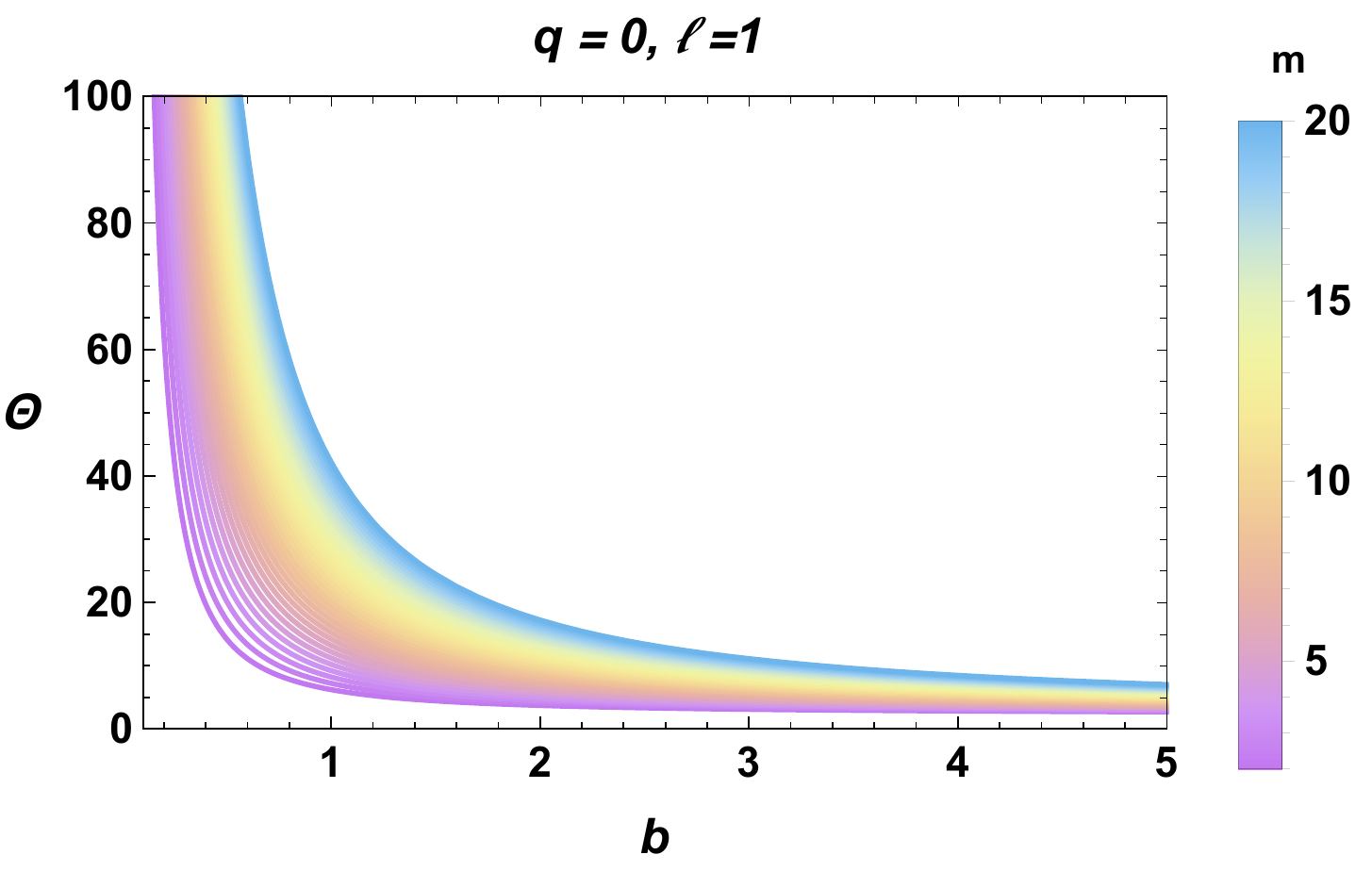} \>
			\includegraphics[scale=.5]{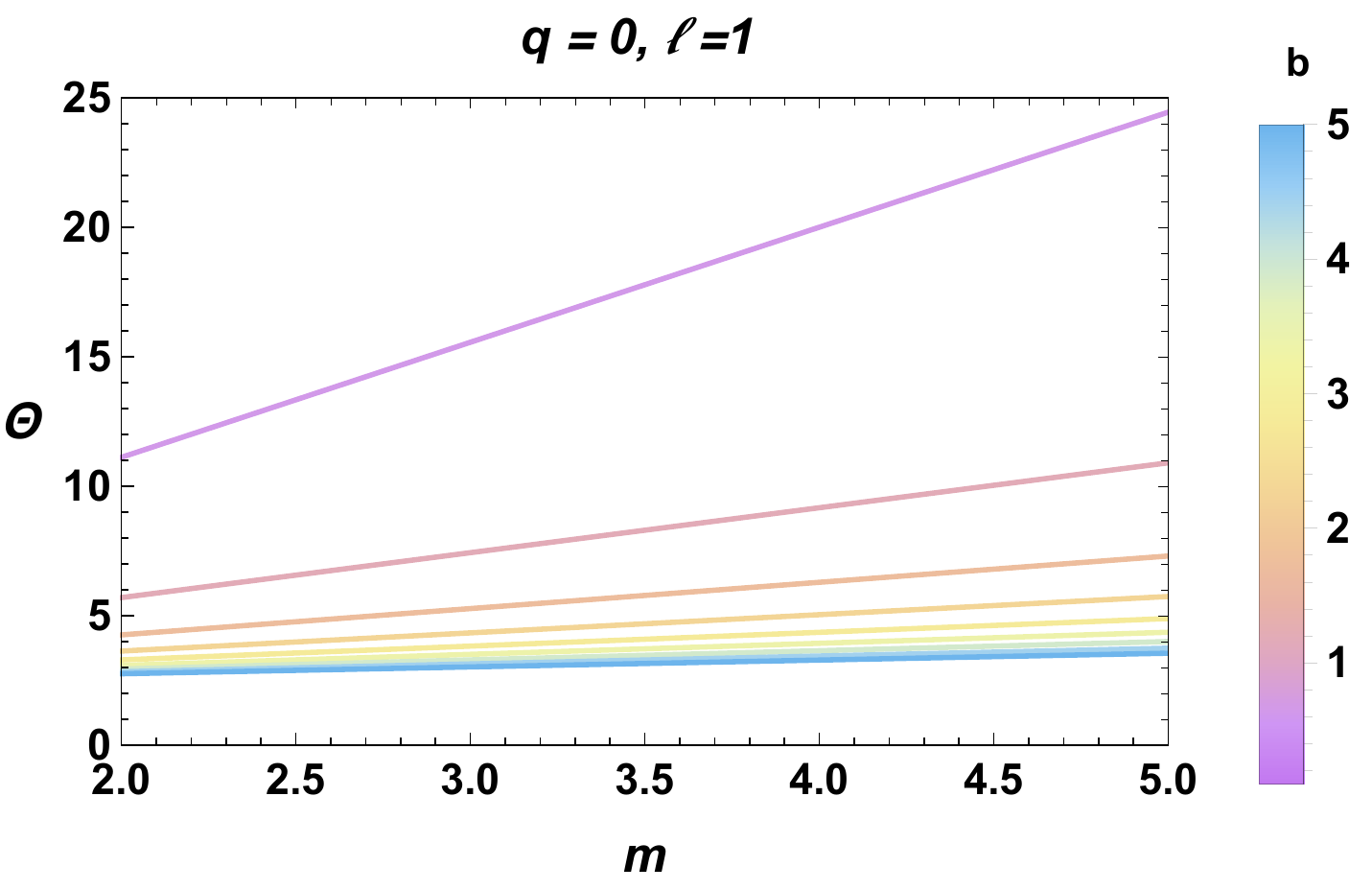} \\

					   \end{tabbing}
					   \vspace{-1 cm}
\caption{{\it \footnotesize 
The deflection angle    behaviors by varying  the  impact parameter  $b$  
 and   the mass $m$ for fixed value of $\ell$}}
\label{shfa2}
\end{center}
\end{figure}

The involved constraint Eq.(\ref{ES})  associated with the building of  such black holes requires   the implementation of the  mass parameter in the investigation of the deflection angle variations.   This way could be considered as an extended  vision   of such an optical quantity.  In particular, we investigate the effect of the mass parameter.   For neutral solutions providing relevant aspects,  the variations are presented in  Fig.(\ref{shfa2}).      
Considering the mass as a parameter,   the left plot shows the variation in terms of the impact parameter.  It has been observed that  the deflection angle increases with the mass parameter.   For large values of $b$, the angle $\Theta$ goes the  previous specific value.   Taking the mass as a variable, the right plot  provides the deflection variations  for different values of $b$ by fixing $\ell$ and $q$. It has been   remarked a linear  variation.  For generic  values of $b$, the deflection angle increases with $m$.   In particular,  it increases  significantly for small values of $b$.  Fixing the mass parameter,  we recover the same result obtained in the previous illustrations.

 \section{Conclusions and final remarks}
 
 In this  paper, we have investigated the weak gravitational lensing in the framework of superentropic black hole solutions.   In particular,  we have considered the photon rays into the equatorial plane. Then,  we have computed the   optical metric  in order to get the corresponding orbit equations.   Using  Gauss-Bonnet theorem, we have   derived the expression of the  total deflection angle   of lights by superentropic black holes.  Theses computations have been based on certain appropriate approximations.  To inspect the behaviors of the angle deflection, many parameters  have been  used to elaborate  graphical representations.   These parameters  have  provided a bounded region in the associated moduli space required by the superenetropic black hole building.   It has been observed that the relevant parameter is the mass pushing one to consider two situations. In the first situation,  we have considered the mass  as a parameter.   For large values of the impact parameter,  it has been   remarked  that  the deflection angle does not go to zero    contrary to  the  ordinary black holes without AdS backgrounds.  We have found a specific value given by $\frac{\pi}{\sqrt{2}}$.   This deflection angle  distinction  could be linked to the presence of AdS radius  and the black hole geometry effects  even for large values of $b$.    This result matches perfectly with the previous  investigations\cite{B110}.
 In the second situation, the mass has been considered as  a variable.   In particular,  it  has been found that the deflection angle involves   linear behaviors.   For  a generic  mass value, this angle increases   by decreasing the   impact paramter.

This work could  open certain new  investigation gates.  It should be also interesting to implement  the   dark sector   effect.  We  could anticipate  that one can  extend   the present work to others  related topics using Gauss-Bonnet theorem results.

\section*{Acknowledgments}
The authors  would like to thank Y. Hassouni,  K. Masmar, M. Oualid, and  M. B. Sedra for discussions on related topics. 
 This work is partially
supported by the ICTP through AF.   

\end{document}